\documentclass[review]{elsarticle}

\usepackage{lineno,hyperref}
\modulolinenumbers[5]

\journal{arXiv.org}

\usepackage{graphicx}    % needed for figures
\usepackage{dcolumn}    % needed for some tables
\usepackage{bm}           % for math
\usepackage{amssymb}   % for math

\usepackage{multirow}% http://ctan.org/pkg/multirow
\usepackage{hhline}% http://ctan.org/pkg/hhline
\usepackage{times}
\usepackage{color}

	 \usepackage{amsmath} 
	 \usepackage{amsthm}
	 \usepackage{bm}
	 \usepackage[mathscr]{eucal}
	 \usepackage{enumerate}

	\usepackage[]{graphicx}
	\usepackage{tabularx}
	\usepackage{array}
	\usepackage{epstopdf}
	\DeclareGraphicsExtensions{.pdf,.eps,.png,.jpg,.mps}
	\usepackage{color}

\usepackage{soul}
\usepackage{float}

\usepackage{tikz}
\newcommand*\circled[1]{\tikz[baseline=(char.base)]{
            \node[shape=circle,draw,inner sep=2pt] (char) {#1};}}
\usepackage{hyperref}
\hypersetup{
 colorlinks,
 citecolor=blue,
 }

\def\##1{{\bf #1}}
\def\=#1{\underline{\underline #1}}

\def\.{\mbox{ \tiny{$^\bullet$} }}

\def\lec{\left\{}
\def\ric{\right\}}

\def\ux{\hat{\#x}}
\def\uy{\hat{\#y}}
\def\uz{\hat{\#z}}

\usepackage{upgreek}
\newcommand{\SImu}{\ensuremath{\upmu}}
\newcommand{\SImum}{\ensuremath{\upmu}\textrm{m}}
\def\MFTF{{\SImu}{\rm FTF}}

\begin{document}
		
 \title{Parylene-C microfibrous  thin films as phononic crystals}
\author
{Chandraprakash Chindam, Akhlesh Lakhtakia$^{\ast}$, and Osama O. Awadelkarim
 \\
	\normalsize{{\it Department of Engineering Science and Mechanics, Pennsylvania State University, 	University Park, PA 16802, USA}}\\
\normalsize{$^\ast$To whom correspondence should be addressed; E-mail: akhlesh@psu.edu}
}
\date{\today}

\begin{abstract}
Phononic bandgaps of Parylene-C  microfibrous thin films  ($\MFTF$s)   were computationally determined by treating them as phononic crystals comprising 
identical  microfibers arranged either on a square or a hexagonal lattice. The microfibers
could be columnar, chevronic, or helical in shape, and  the host medium could
be either  water or air.  All   bandgaps were observed to lie in the 0.01--162.9-MHz regime,
for microfibers of realistically chosen dimensions. The upper limit of the frequency of bandgaps was the highest
for the columnar $\MFTF$ and the lowest for the chiral $\MFTF$. More bandgaps
exist when the host medium is water than air.
Complete bandgaps were observed for the columnar $\MFTF$ with microfibers arranged on a hexagonal lattice in air, the
 chevronic $\MFTF$ with microfibers arranged on a square lattice in water, and 
the chiral $\MFTF$ {with microfibers arranged on a hexagonal lattice in either air or water}. 
The softness of the Parylene-C $\MFTF$s  makes them mechanically tunable,
and their bandgaps can be exploited in
multiband  ultrasonic  filters.
 \end{abstract}


\maketitle

\section{Introduction}

Products are generally designed to accomplish only one specific function \cite{product_design}. 
But, multifunctionality~---~the  ability to perform multiple functions~---~is currently emerging as a technoscientific  paradigm inspired by nature \cite{Nicole, multifun,mimume}. 
Two prime examples of natural multifunctional entities \cite{Matic} are the leaves of  plants and the skins of animals.
 Leaves are designed to transport nutrients across the neighboring cells, release carbon dioxide, and  absorb sunlight, and some  are capable of self-cleaning \cite{Self_clean}. 
Skins of animals, including humans,  hold the body parts in place and maintain the shape
of the body, facilitate perspiration, hold hair  for thermal insulation, and also provide the sense of touch. For a sustainable future \cite{sustainability} we need to decrease the use of single-function devices by devising and popularizing multifunctional devices. 
Some engineered products perform multiple functions; for example, printers copy, print, staple, email, and  fax. Although such products usually are conglomerations of many single-function devices,  a few dual-function devices and structures have been reported. Examples include light-emitting diodes that also function as
photodetectors \cite{dual_function}, and  photonic-cum-phononic crystals \cite{PhX2}.

In the current age of miniaturization, many devices contain thin films for which fabrication techniques are well-known \cite{Donald_Smith, maldovan2009periodic, STFbook,Piegari}. 
Thin films can be so highly dense as to be considered homogeneous \cite{Donald_Smith,Piegari} or can be porous with engineered morphology \cite{STFbook,Piegari}.
Microfibrous thin films  ({\MFTF}s) of the polymer Parylene~C are attractive as multifunctional materials \cite{mimume}. The Parylene-C {\MFTF}s are fabricated by a straightforward modification \cite{PHDL,LaiWei_12_1_MRI}
of the   industrially used Gorham process \cite{Gorham}  to coat various structures conformally with dense Parylene-C films \cite{Licari,Pereira,Vernekar}. Thus far,
Parylene-C $\MFTF$s  comprising parallel and identical microfibers of   upright circular-cylindrical, slanted-circular cylindrical, chevronic,  and helical shapes have been fabricated \cite{WLRR}. The water-wettability \cite{Chan_AppSci}, crystallinity \cite{Chan_AppSci}, ability to store electric charge \cite{Ibrahim_ECS_2015},   glass-transition temperature \cite{Chan_Mech}, and biocompatibility\cite{WLRR,protein_assay}   of Parylene-C $\MFTF$s comprising slanted-circular cylindrical microfibers have been experimentally investigated.

If the identical microfibers are deposited on a topographic substrate decorated with a
regular lattice \cite{Horn03_1Nanotechnology,Dutta}, the Parylene-C $\MFTF$ could function
as a phononic crystal \cite{Pennec2010229,Deymier}. Phononic crystals are being investigated as acoustic sensors \cite{sensing,sensing2},  isolators \cite{Sandia}, 
filters \cite{Golub,Wilson},  waveguides \cite{Wilson}, and concentrators \cite{TTWu}.
Most studies consider an array of solid/fluid  scatterers in a fluid/solid medium, the scatterers being of simple shapes such as infinitely long cylinders \cite{Pennec2010229} and spheres \cite{sphere_PWE}.
Also in these studies, the  scatterers are arranged on either a
square or a hexagonal lattice in two-dimensional  space, or in either a face- or a body-centered  cubic lattice in three-dimensional space.
 
As we were interested in investigating the $\MFTF$s of Parylene~C as ultrasonic filters,
we determined their phononic-bandgap characteristics using the commercial finite-element-method (FEM) software COMSOL Multiphysics$^{\rm \textregistered}$ (version 5.1) \cite{Comsol}. We consider all microfibers in a $\MFTF$ to be identical and parallel. $\MFTF$s comprising  circular-cylindrical  microfibers are referred to as columnar $\MFTF$, and those  with chevronic and helical microfibers as chevronic and chiral $\MFTF$s, respectively. 
As  Parylene-C $\MFTF$s are easily removable from topographic substrates \cite{LaiWei_12_1_MRI},  we consider the free-standing  $\MFTF$ as a periodic arrangement of microfibers with the host medium as air, i.e., a phononic crystal with host as air.
Furthermore, we also examine the phononic dispersion characteristics of the $\MFTF$ in water keeping biomedical research in mind \cite{WLRR,protein_assay}. As Parylene~C is a polymer rather than a hard material such as a metal, the Parylene-C $\MFTF$s are very different from most phononic crystals reported in the literature \cite{Pennec2010229,Deymier,sensing,sensing2,Sandia,Golub,Wilson,TTWu,sphere_PWE} in that the Parylene-C $\MFTF$s shall be tunable by the application of pressure
\cite{FeiWang} in the same way that liquid-crystal elastomers are \cite{SKF,Shibaev,Varanytsia}. 

The plan of this paper is as follows. 
We   describe the geometric details of the chosen $\MFTF$s in Sec.~\ref{GtP} and theoretical basis (governing constitutive relations and Brillouin zone paths)  used for determining the eigenfrequencies in Sec.~\ref{ThP}.
The procedure to implement COMSOL Multiphysics$^{\rm \textregistered}$~
and the validation of that procedure are briefly described in Sec.~\ref{comsol_details}. 
Finally in Sec.~\ref{RD} we present the results of the phononic bandgaps of Parylene-C $\MFTF$s. 
A time  dependence  of $\exp(i{\omega}t)$  is implicit, with $i=\sqrt{-1}$, $t$ as time,  $\omega=2\pi{f}$ as the angular frequency, and $f$ as the linear frequency. 
The microfibers are arranged on either a square or a hexagonal lattice in the $xy$ plane.
Vectors are denoted in boldface,   every unit vector is identified by a caret, and $\#r =x\ux + y\uy+z\uz$ denotes the position vector. 
 
\section{Geometric Preliminaries}\label{GtP}

\begin{figure}[H]
			\begin{center}
				\includegraphics[width=3.4in]{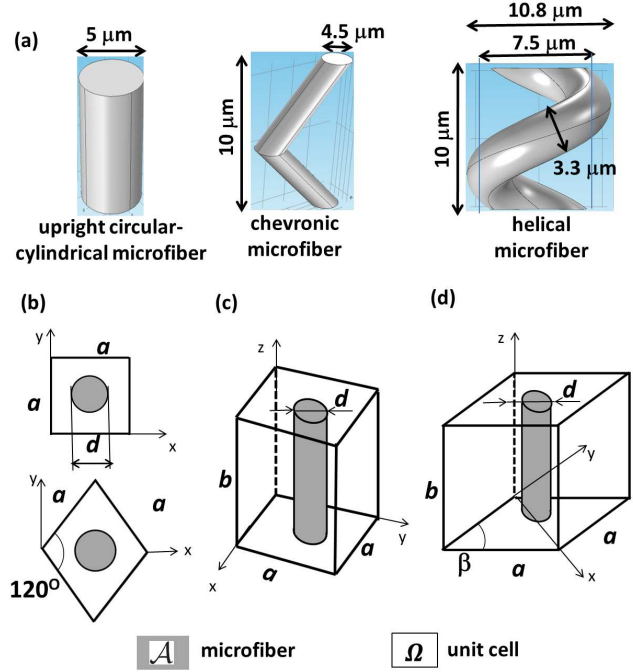}
			\caption{(Color online) (a) Microfibers of  upright circular-cylindrical, chevronic, and structurally right-handed helical  shapes with their  dimensions chosen for numerical results presented here.  (b) Square and hexagonal lattices used for the columnar $\MFTF$. Unit cells of (c) square and (d) hexagonal lattices   used for the chevronic and chiral $\MFTF$s, with lattice dimensions $a$ and $b$, the microfiber diameter denoted by $d$.  In (d), the angle $\beta=2\pi/3$ for chevronic $\MFTF$s and $\beta=\pi/3$ for chiral $\MFTF$s. In (b)--(d), the unit cell is denoted by $\Omega$, the shaded region $\cal A$ is completely occupied by the Parylene-C microfiber, and the unshaded region $\Omega-\cal A$ by either air or water.
			{For the square lattice, the chosen $\MFTF$s with upright circular-cylindrical, chevronic, and helical microfibers have filling fractions of  0.20, 0.16, and 0.15, respectively; for the hexagonal lattice, the corresponding filling fractions
			are 0.23, 0.18, and 0.11, respectively.}
} 
			\label{Unit_cells}
			\end{center}
			\end{figure}

In a preceding study \cite{Chan_AS_JAP}, we investigated  the scattering of an acoustic 
plane wave by a single finite-sized microfiber of Parylene C in water.
 The microfiber could
be upright circular-cylindrical, slanted-circular cylindrical, chevronic, or helical in shape.
However, as a phononic crystal is considered to occupy all space for the purpose of
determining its phononic dispersion characteristics, only a
simple rotation of the coordinate system is needed to see that a $\MFTF$ comprising  slanted circular-cylindrical microfibers is identical
to a  $\MFTF$ comprising upright circular-cylindrical microfibers. Hence, 
Parylene C $\MFTF$s comprising upright circular-cylindrical, chevronic, and helical microfibers, shown in Fig.~\ref{Unit_cells}(a),  were chosen for investigation as phononic crystals.

%%%%%%%% Figure 1 begins\ %%%%%%%%%%%%%%%%%%%%%
%%%%%%%% Figure 1 ends %%%%%%%%%%%%%%%%%%%%%%

The upright circular-cylindrical microfiber is parallel  to the $z$ axis and of infinite length, so that the lattice is really two-dimensional, as shown
 in Fig.~\ref{Unit_cells}(b). 
This lattice can be either square or hexagonal, the lattice dimension being denoted by $a$.
 The chevronic and chiral microfibers are periodic along the $z$ axis with period $b$, the three-dimensional unit cells drawn on square and hexagonal lattices of dimension $a$, as shown in Figs.~\ref{Unit_cells}(c,d). 
The chevron underlying the chevronic microfiber lies in the $xz$ plane. The helix underlying the chiral microfiber is curled about the $z$ axis, the helix being either left- or right-handed. We took the chiral microfibers to be right-handed for all calculations reported here.  The microfiber diameter is denoted by $d$ in Figs.~\ref{Unit_cells}(b)--(d).

The microfiber dimensions shown in Fig.~\ref{Unit_cells}(a) were chosen as representative values, based on  several  scanning-electron micrographs of Parylene-C $\MFTF$s \cite{WLRR,Chan_AppSci, protein_assay,Chan_AS_JAP}. The  distance in the $xy$ plane  between a microfiber and its nearest neighbor
 distance can be  chosen above a minimum value  by topographically patterning the substrate \cite{LaiWei_12_1_MRI,Horn03_1Nanotechnology}, but for all calculations reported here
 the lattice parameters were fixed as follows. The square lattice was
 chosen of side $a=10~\SImum$ and the hexagonal lattice of side $a= 10~\SImum$
 in Figs.~\ref{Unit_cells},
 when the  microfibers are either columnar or chevronic. The square lattice was
 chosen of side $a=12~\SImum$ and the hexagonal lattice of side $a= 15~\SImum$,
 when the  microfibers are helical.
As appropriate, the dimension $b = 10~\SImum$ was fixed.

Since the the inter-microfiber space in as-fabricated $\MFTF$s is occupied by air,  we studied the phononic crystals with air as the host medium. Keeping their biomedical applications in mind \cite{WLRR,protein_assay}, we also studied the phononic crystals with water as the host medium.

\section{Theoretical Preliminaries}\label{ThP}

 The unit cell $\Omega$ is two-dimensional for columnar $\MFTF$s (Fig.~\ref{Unit_cells}(b)), but three-dimensional for chevronic $\MFTF$s and chiral $\MFTF$s (Figs.~\ref{Unit_cells}(c--d)).
The unit cell is also the domain of all  calculations for phononic crystals. The displacement phasor $\#u(\#r)$ in an isotropic material  is governed by
 the Navier  equation \cite{Hetnarski}
\begin{equation}
[\lambda(\#r) + \mu(\#r) ] \nabla \left[\nabla \. \#u(\#r)\right] + \mu(\#r) \nabla^2\#u(\#r) + \rho(\#r)\omega^2\#u(\#r)=\#0\,,
\label{Nav_eq}
\end{equation} 
where  the Lam\'e parameters $\lambda(\#r)$ and $\mu(\#r)$ as well as the mass
density $\rho(\#r)$ are piecewise continuous as follows:
\begin{eqnarray}
\nonumber &&
\lambda(\#r)=\left\{\begin{array}{l}
\lambda_s\\
\lambda_h
\end{array}\right.,\quad
\mu(\#r)=\left\{\begin{array}{l}
\mu_s\\
\mu_h
\end{array}\right.,\quad
\\
&&
\rho(\#r)=\left\{\begin{array}{l}
\rho_s\\
\rho_h
\end{array}\right.,\quad
\#r\in \left\{\begin{array}{l}
{\cal A}
\\
\Omega-{\cal A}
\end{array}\right.\,.
\end{eqnarray}

Both traction and displacement were set to be continuous across every interface. 
The Floquet--Bloch periodicity condition \cite{Hussein2825} was imposed,  as appropriate, on the boundaries of the unit cell.  With $\#k = k_x\ux+k_y\uy+k_z\uz$ denoting the wave vector, 
the displacement, the mass density, and the Lam\'e parameters
are expanded as the Fourier series
\begin{equation}
\left.\begin{array}{l}
\#u(\#r) = \exp(-i\#k\. \#r)\sum_{\#G}\,\left[\#u_{\#G}\exp\left(-i 2\pi  \#G  \.\#r\right)\right]\\[4pt]
\rho(\#r) = \sum_{\#G}\,\left[\rho_{\#G}\exp(-i2\pi\#G\. \#r)\right]\\[4pt]
\lambda(\#r)= \sum_{\#G}\,\left[\lambda_{\#G}\exp(-i2\pi\#G\. \#r)\right]\\[4pt]
\mu(\#r)= \sum_{\#G}\,\left[\mu_{\#G}\exp(-i2\pi\#G\. \#r)\right]
\end{array}\right\}\,,
\label{expansions}
\end{equation}
where  the Fourier coefficients $\rho_{\#G}$, $\lambda_{\#G}$, and $\mu_{\#G}$ are known, but $\#u_{\#G}$
is unknown for all
$\#G = n_1{\#b_1}+n_2{\#b_2}+n_3{\#b_3}$. Here, $\left\{\#b_1,\#b_2,\#b_3\right\}$ is
the triad of basis vectors of the reciprocal lattice space, whereas the integers
$n_\ell\in   (- \infty, \infty )$ for all $\ell\in\left\{1,2,3\right\}$.
 If
$\left\{\#a_1,\#a_2,\#a_3\right\}$ is the triad of basis vectors of the lattice,
then the vectors ${\#b_1} =  { ({\#a_2}\times {\#a_3})}/{V_\Omega}$, ${\#b_2} =  { ({\#a_3}\times {\#a_1})}/{V_\Omega}$, and $ {\#b_3} =  { ({\#a_1}\times {\#a_2})}/ V_\Omega$, whereas
  ${V_\Omega} = {{\#a_1}\.({\#a_2}\times {\#a_3})} $ is the volume of $\Omega$. 
 The Fourier coefficients $\gamma_{\#G} $ of   $\gamma \in\lec \rho,\lambda,\mu \ric$  are given by \cite{Kushwaha}
\begin{equation}
\gamma_{\#G}=  \frac{1}{V_\Omega}\int_{\Omega}\gamma(\#r)\exp(i2\pi\#G\. \#r)\,d\#r\,.
\end{equation}
 After noting that  
\begin{equation}
\int_{\Omega} \exp[i2\pi(\#G-\#G^\prime)\.\#r] \,d\#r = V_\Omega\,\delta(\#G-\#G^\prime)
\,,
\end{equation}
where $\delta(\#G-\#G^\prime)$ is the Dirac delta,
we get 
\begin{equation}
\gamma_{\#G} =
  \begin{cases}
    \gamma_{s} p+ \gamma_h (1-p)\,,  &  \#G\neq \#0 \\[5pt]
   (\gamma_{s}- \gamma_h)P({\#G}) \,,      &  \#G \neq \#0
  \end{cases}
  \label{detail_exp}
\end{equation}
where the filling fraction $p$ is the ratio of the volume $V_{\cal A}$ of $\cal A$ to  $V_\Omega$ and the  structure function 
\begin{equation}
P(\#G)=  \frac{1}{V_\Omega}
\int_{\cal A}\exp(i2\pi\#G \. \#r) d\#r\,.
\end{equation}

In the planewave expansion method (PWEM) \cite{sphere_PWE}, substitution of Eqs.~(\ref{expansions}) in Eq.~(\ref{Nav_eq}), followed by restricting
$n_\ell\in  [- N, N  ]$ for all $\ell\in\left\{1,2,3\right\}$ with a sufficiently large
natural number  
$N$  for computational tractability, yields a set of linear algebraic equations for $\#u_{\#G}$.
This set can be be cast as a matrix eigenvalue  problem \cite{sphere_PWE}. 
The eigenvalues (i.e., the eigenfrequencies) can be evaluated using MATLAB$^{\rm \textregistered}$ \cite{matlab} (version R2012a) when both the microfiber (or any other scatterer) and the host medium are solids. However, if either the scatterer medium or the host medium
is a fluid, then either $\mu_s=0$  or $\mu_h=0$, resulting in a large sparse matrix which is ill conditioned \cite{Singularity}. As $\mu_h=0$ for our problem, as noted in Sec.~\ref{GtP}, the PWEM could not be used, and we used the commercial FEM software
COMSOL  Multiphysics$^{\rm \textregistered}$.

%%%%%%%% Figure 2 begins\ %%%%%%%%%%%%%%%%%%%%%
	\begin{figure}[H]
			\begin{center}
				\includegraphics[width=3.4in]{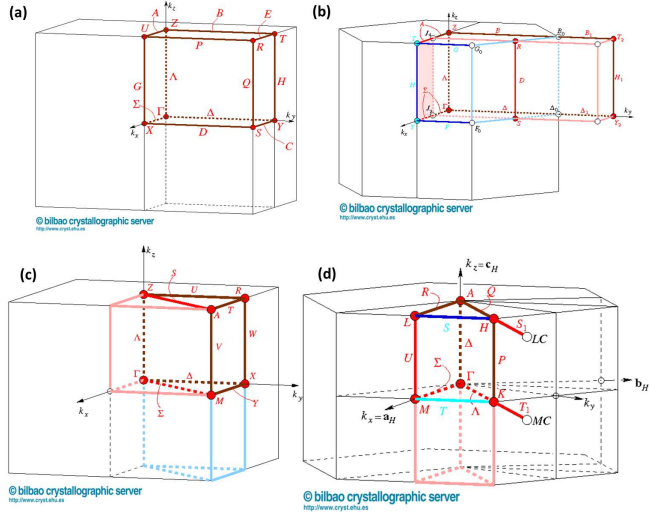}
			\caption{(Color online) Irreducible brillouin zones for the chevronic microfiber arranged in  (a) square and (b) hexagonal lattices \cite{chevron_sc}. (c) and (d) are the same as (a) and (b) but for a  helical microfiber \cite{chiral_sc}. [By kind permission of Dr. Mois Ilia Aroyo on behalf of the Bilbao Crystallographic Server.] } 
			\label{IBZs}
			\end{center}
			\end{figure}
%%%%%%%% Figure 2 ends %%%%%%%%%%%%%%%%%%%%%%

For the square lattice regardless of the microfiber shape,  ${\#a_1} = a \ux$, ${\#a_2} = a\uy$, and ${\#a_3} = b \uz$; 
for the  hexagonal lattice  for  columnar and chiral $\MFTF$s,  ${\#a_1} = a\left(\ux +\sqrt{3}\uy\right)/2 $, ${\#a_2} = a (\ux -\sqrt{3}\uy)/2 $, and ${\#a_3} = b \uz$; and for the
hexagonal lattice for chevronic $\MFTF$s with the microfibers oriented to lie in the $xz$
plane, ${\#a_1} = a(\sqrt{3}\ux +\uy)/2 $, ${\#a_2} = a (\sqrt{3}\ux -\uy)/2 $, and ${\#a_3} = b \uz$.

The irreducible Brillouin zone (IBZ) and the coordinates of the corners for the three repeating units of microfibers arranged as square and hexagonal lattice  \cite{chevron_sc, chiral_sc} were taken from the Bilbao Crystallographic Server \cite{bilbao_2, bilbao_1}.  
The corresponding IBZs are shown in Fig.~\ref{IBZs}.
Eigenfrequencies were estimated for each path of the IBZ separately by setting the values of the components of the wave vector as different lengths of the paths of IBZ. 
These eigenfrequencies were assembled  in MATLAB$^{\rm \textregistered}$ to get the complete band diagrams.

%%%%%%%%%%%%%%%%%%%%%%%%%%%%%%%%%%%%%%%%%%%%%

\section{Methodology}\label{comsol_details}

\subsection{Implementation for $\MFTF$s}\label{mesh_details}
The domain and boundary conditions mentioned in Sec.~\ref{ThP} were  implemented in COMSOL Multiphysics$^{\rm \textregistered}$, for the unit cells  shown in Fig.~\ref{Unit_cells}.
We imposed the  Floquet--Bloch periodic conditions  by setting $\#k = \#{k}_F$, where $\#k_F =2\pi( f_1\#b_1+f_2\#b_2+ f_3\#b_3)$,  with $\{f_1,f_2\} \in [-1/3, 2/3]$ and $f_3 \in [0 , 1/2]$.

  %%%%%%%% Figure 3 begins\ %%%%%%%%%%%%%%%%%%%%%
	\begin{figure}[H]
			\begin{center}
				\includegraphics[width=3.4in]{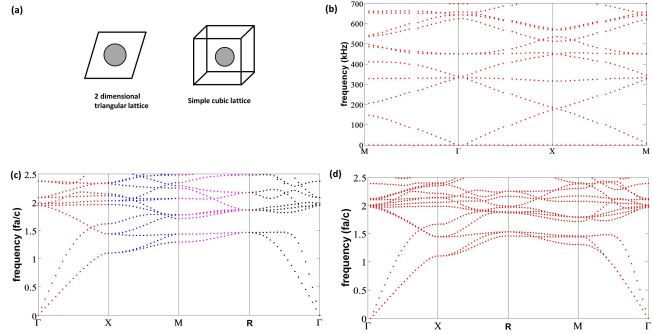}
			\caption{(Color online) Meshed   unit cells. } 
			\label{Mesh_image}
			\end{center}
			\end{figure}
%%%%%%%% Figure 3 ends %%%%%%%%%%%%%%%%%%%%%%

Fig.~\ref{Mesh_image} shows  images of all unit cells meshed for implementing the finite-element method on the chosen $\MFTF$s.
The following meshing procedures  were used to generate the meshes. For the two-dimensional lattice used for the columnar $\MFTF$, a {\it Predefined: Fine} mesh mode was set in both $\cal A$ and $\Omega-{\cal A}$, with the maximum and minimum element sizes  set as $0.53~\SImum$ and $0.003~\SImum$, respectively, along with a `maximum element growth rate\rq{} of 1.3. 
One pair of adjacent edges of the square were meshed and copied to the opposite edges to mesh the boundaries. 	
Following that step,  a `Free Triangular\rq{} mesh was set  in both $\cal A$ and $\Omega-{\cal A}$.

For the unit cells of the chevronic and chiral  $\MFTF$s,
the  \lq Mapped mesh\rq~was set on two faces whose  normals are ${\#a_1} \times {\#a_3}$ and ${\#a_2} \times {\#a_3}$,  and a \lq distribution\rq~of 12 elements was set on all the edges of these two faces.  
 Next, the mapped mesh on these faces was copied to the opposite faces. 
 A \lq Free Quad\rq~mesh was set on the  faces parallel to which  ${\#a_1}\times {\#a_2}$ is normal,  on both the microfiber and host.
 All these meshed faces were \lq Converted\rq~ by the \lq Element Split method\rq~via  the {\it Insert diagonal edges} option. 
Thereafter, the `quad mesh\rq{} on the bottom face was copied to the opposite face and a \lq Free Tetrahedral\rq~mesh was set within the  entire domain. 
{For all  unit cells, initial checks were made with every microfiber to verify the sufficiency of  \emph{Predefined: Fine} mesh over \emph{Predefined: Finer} mesh. With either kind of  mesh, the convergence value was less than $10^{-6}.$ As the computation time was faster for a \emph{Predefined: Fine} mesh than for the \emph{Predefined: Finer} mesh, we chose the former for all calculations.}

The `parametric sweep\rq{} option was used to determine the set of eigenfrequencies  of every  unit cell  for different choices of $\#{k}_F$. 
For every  $\#{k}_F$, 20 eigenfrequencies  were determined with a \lq search method around shift\rq{} as {\it larger real part} around $10^6$~Hz.
Based on a perusal of the literature, this constraint is
in accord with the conjecture that any eigenfrequency
 in a bandgap is likely to fall in the range $[0, 2c_h/a]$,where  $c_h \equiv \sqrt{(\lambda_h+2\mu_h)/\rho_h}$ is the longitudinal-wave speed in $\Omega-\cal A$. This constraint also determined the highest frequency for each of the band diagrams presented in this paper.
 
 In the `solver configuration\rq{} option, the \lq search method around shift\rq~was set as \lq closest in absolute value\rq~option with the \lq transformation value\rq~as $10^6$.
 The MUltifrontal Massively Parallel Sparse (MUMPS) \cite{mumps} direct solver provided in COMSOL  was used with a memory allocation factor of $1.2$. 

\subsection{Validation of  methodology}

\begin{figure}
			\begin{center}
				\includegraphics[width=3.4in]{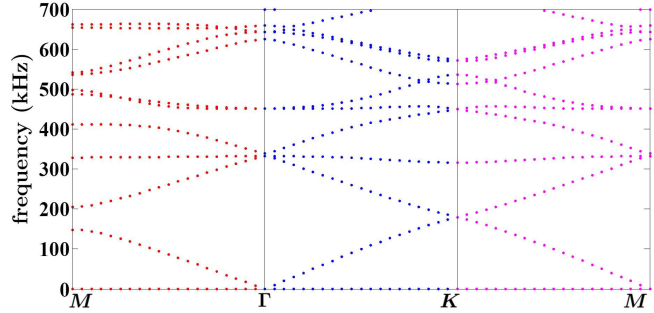}
			\caption{(Color online)  Band diagram calculated using the
			commercial  FEM  software COMSOL Multiphysics$^{\rm \textregistered}$
			 for a hexagonal array ($a=4.5$~mm) of
			infinitely long circular cylinders (radius $R= 1.4143$~mm) of steel in water. This diagram is identical to a published result \cite[Fig.~2(b)]{hexagonal}. } 
			\label{Validation1}
			\end{center}
			\end{figure}

To validate the numerical procedure implemented in COMSOL Multiphysics$^{\rm \textregistered}$, the  two  check cases  were chosen. The first check case is an array of infinitely long, parallel, solid circular cylinders 
of radius  $R=1.4143$~mm arranged on a two-dimensional  hexagonal lattice
($a=4.5$~mm) in  a fluid host medium \cite{hexagonal}.
 Each cylinder
was taken to be made of steel ($\lambda_s =  148.96$~GPa, $\mu_s=73.37$~GPa,   $\rho_s=7890$~kg~m$^{-3}$) and host medium was water
($\lambda_h =2.19$~GPa, $\mu_h=0$,  $\rho_h=1000$~kg~m$^{-3}$). 
With ${\#a_1} = a\left(\ux +\sqrt{3}\uy\right)/2 $ and ${\#a_2} = a \left(\ux -\sqrt{3}\uy\right)/2 $, we obtained  $\#b_1=(2/3a^2)
(2\#a_1+\#a_2)$ and
$\#b_2=(2/3a^2)
(\#a_1+2\#a_2)$.
The  IBZ is the path connecting the points 
$\Gamma$~(i.e., $\#G=\#0$), 
$M$~(i.e., $\#G={\#b_2}/2$), and $X$~(i.e., $\#G={\#b_1}/3 \#+  {\#b_2}/3$).
The mesh    for this unit cell was chosen following the procedure described for  columnar $\MFTF$s  in Sec.~\ref{mesh_details}, and the eigenfrequencies of the unit cell were determined.
Figure~\ref{Validation1}  shows the band diagram calculated by us in this way,
every band  represented by a dotted line that spans a specific path of the IBZ.
The band diagram in Fig.~\ref{Validation1}
is identical to a published result  \cite[Fig.~2(b)]{hexagonal}.

%%%%%%%% Figure 4 begins\ %%%%%%%%%%%%%%%%%%%%%

%%%%%%%% Figure 4 ends %%%%%%%%%%%%%%%%%%%%%%

For the second check case, solid spheres ($\lambda_s=101.92$~GPa, $\mu_s= 80.08$~GPa,  $\rho_s=7850$~kg~m$^{-3}$) of radius $R=3~\SImum$ arranged in
 a simple cubic lattice of side $a = 10~\SImum$ in a solid host medium
 ($\lambda_h= 458.05$~GPa, $\mu_h= 147.99$~GPa,  $\rho_h=1142$~kg~m$^{-3}$) were considered. Since  ${\#a_1} = a \ux$, ${\#a_2} = a \uy$, and ${\#a_3} = a \uz$ for the
 simple cubic lattice, we get ${\#b_1} = a^{-1} \ux$, ${\#b_2} = a^{-1}\uy$,   ${\#b_3} = a^{-1}\uz$, $p = (4\pi/3) (R/a)^3$, and
\begin{equation}
P({\#G}) = \frac{3p}{({GR})^3}[\sin( {GR})-{GR}\cos( {GR})]\,,
\end{equation}
where $G=\vert\#G\vert$.
The corresponding IBZ is the path connecting 
$\Gamma$~(i.e., $\#G=\#0$), 
$X$~(i.e., $\#G={\#b_1}/2$), 
$M$~(i.e., $\#G={\#b_1}/2\#+ {\#b_2}/2$), and 
$R$~(i.e., $\#G={\#b_1}/2\#+{\#b_2}/2\#+ {\#b_3}/2$). 

\begin{figure}
			\begin{center}
				\includegraphics[width=3.4in]{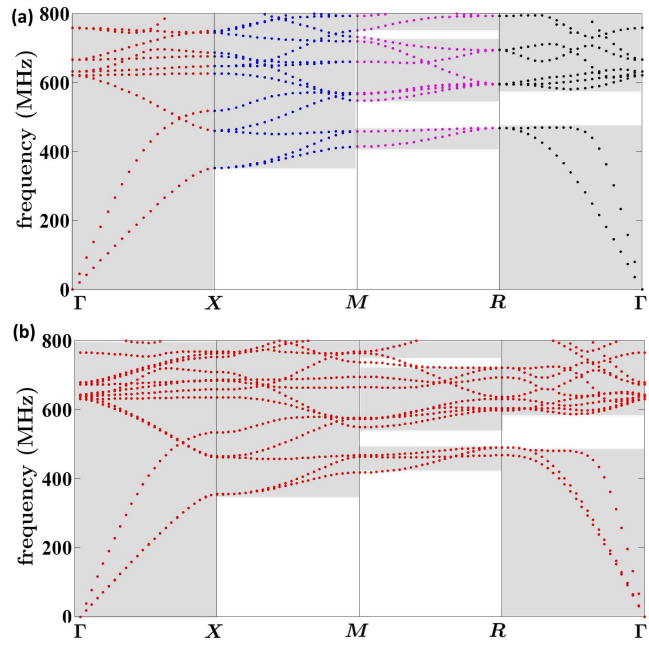}
			\caption{(Color online) Comparison of band diagrams, for solid spheres  of radius $R=3~\SImum$ arranged on a cubic lattice ($a = 10~\SImum$) in a solid host medium, obtained using the (a)  FEM and (b) PWEM.
Note that $\lambda_s=101.92$~GPa, $\mu_s= 80.08$~GPa,   $\rho_s=7850$~kg~m$^{-3}$,
$\lambda_h= 458.05$~GPa, $\mu_h= 147.99$~GPa,  and $\rho_h =1142$~kg~m$^{-3}$. }
			\label{Validation2}
			\end{center}
			\end{figure}

As the
 sphere in the unit cell does not touch the   boundary of the unit cell,  the following meshing procedure was  used for FEM implemented in COMSOL Multiphysics$^{\rm \textregistered}$. 
A \lq user-controlled mesh\rq~of {\it fine} quality was set for the entire geometry. 
A \lq Mapped mesh\rq~was set on three faces whose normals  are  ${\#a_1}$, ${\#a_2}$, and ${\#a_3}$    and a  \lq distribution\rq~of 12 elements on all the edges of these three faces. 
 All these meshed faces were \lq Converted\rq~ by the \lq Element Split method\rq~via \lq Insert diagonal edges,\rq  
 a \lq Free Tetrahedral\rq~mesh was set in $\Omega$,
  and the eigenfrequencies were evaluated over the IBZ. Also, for comparison, the PWEM   \cite{sphere_PWE},  described in Sec.~\ref{ThP}, was  implemented using MATLAB$^{\rm \textregistered}$  to determine the eigenfrequencies of the unit cell.
 Figs.~\ref{Validation2}(a) and (b), respectively, show the band diagrams obtained
 using  the FEM and PWEM. Despite some differences,
we  found very good agreement between the two band diagrams, which thus provided confidence in
our  COMSOL Multiphysics$^{\rm \textregistered}$ implementation. 
 
%%%%%%%% Figure 5 begins\ %%%%%%%%%%%%%%%%%%%%%

%%%%%%%% Figure 5 ends %%%%%%%%%%%%%%%%%%%%%%

\section{Results and Discussion}\label{RD}
Let us now present band diagrams for the  Parylene-C $\MFTF$s of each of the three morphologies 
in Fig.~\ref{Unit_cells} calculated using
the following constitutive parameters:
\begin{eqnarray}
\nonumber
&&
{\lambda=\left\{\begin{array}{ll}
3.943 &{\rm GPa} \\
2.2 & {\rm GPa}\\
142 & {\rm kPa}
\end{array}\right.,\quad
\mu=\left\{\begin{array}{ll}
0.986 &{\rm GPa} \\
0\\
0
\end{array}\right.,}
\\
&&\quad
{\rho=\left\{\begin{array}{ll}
1289 &{\rm kg~m}^{-3} \\
1000 & {\rm kg~m}^{-3}\\
1.225 & {\rm kg~m}^{-3}
\end{array}\right.,\quad
\left\{\begin{array}{l}
{\rm Parylene~C}\\
{\rm water}\\
{\rm air}
\end{array}\right.}\,,
\end{eqnarray}
The viscoelastic parameters  of Parylene~C were ignored,    as they are not known at high frequencies \cite{Chan_AS_JAP}.   Apart from a trivial eigenfrequency of 0~Hz (at $\Gamma$
in band diagrams), all eigenfrequencies found exceeded 0.01~MHz.

\subsection{Columnar $\MFTF$}\label{res-col}
\begin{figure}[H]
			\begin{center}
				\includegraphics[width=3.4in]{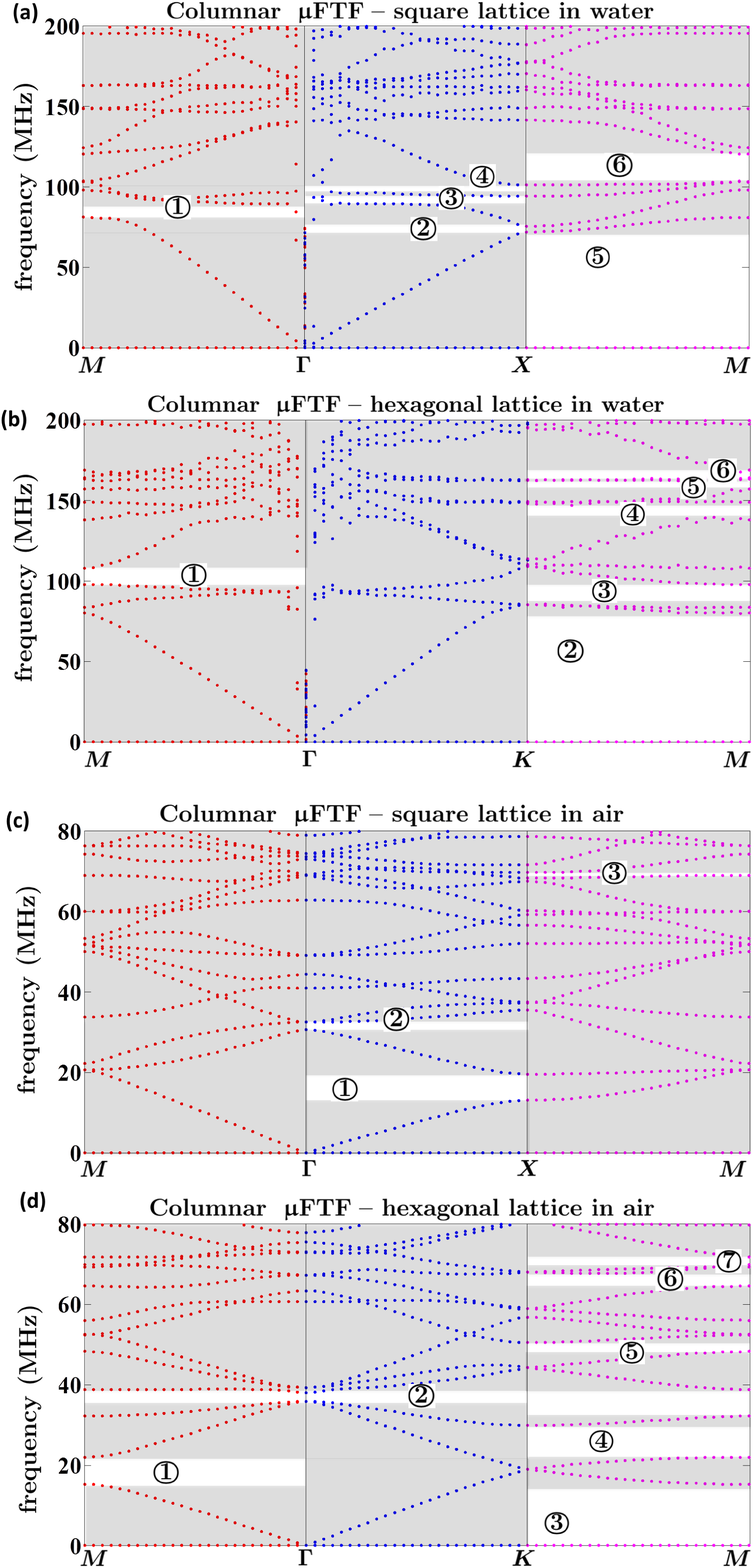}
			\caption{(Color online) Band diagrams for the chosen columnar $\MFTF$, with the upright circular-cylindrical microfibers
			of Parylene~C arranged on either (a,c) a square or  (b,d) a hexagonal lattice,
			the host medium being either (a,b) water or (c,d) air. Bandgaps are shown unshaded   and each is identified by a number inside a circle.}
			\label{PC_circle}
			\end{center}
			\end{figure}

The IBZ of a 2D  unit cell for columnar $\MFTF$s,  with the upright circular-cylindrical microfibers of Parylene~C arranged on either  a square or   a hexagonal lattice has a total of three paths.  The band diagrams for the chosen columnar $\MFTF$ with the host medium being either water or   air are shown in Figs.~\ref{PC_circle}(a)--(d).  

When the host medium is water, we found 6 partial bandgaps  for the square lattice in the   $[0.5,120.7]$~MHz range  and   6 partial bandgaps for the hexagonal lattice   in the $[1.6,162.9]$~MHz
range,
but no complete bandgap is evident in Figs.~\ref{PC_circle}(a,b).
When the host medium is air, we found 3 partial bandgaps  for the square lattice in the  $[0.6, 69.8]$~MHz range and   7 partial bandgaps for the hexagonal lattice   in the $[0.7,71.8]$ MHz range, in addition to a complete bandgap (35.8 to 38.2~MHz) for the hexagonal lattice, in
Figs.~\ref{PC_circle}(c,d). 
The complete bandgap comprises three partial bandgaps.
 The range of
 bandgaps of the  columnar $\MFTF$ immersed in water is about  twice that of  the columnar $\MFTF$  with air as the host medium. 
Regardless of the host medium, the  columnar $\MFTF$ with the hexagonal lattice appears  suitable as a multiple-bandstop filter for the 
 path $KM$  of the IBZ.

A bandgap comprises those frequencies  for which a plane wave cannot pass through the medium of interest.
Since the eigenfrequency calculations were made by assuming that all mediums  in the phononic crystal are
nondissipative, one can expect maximums in the back-scattering efficiencies  $Q_b$ of   a solitary unit cell in the bandgaps. 
In a predecessor study  on the scattering characteristics of an
individual circular-cylindrical microfiber of Parylene~C  (of the same dimensions as in a
unit cell) and 
immersed in water \cite[Fig.~8(a)]{Chan_AS_JAP}, we observed peaks in the spectrum of $Q_b$  at 106, 116, 133, 146, 165, 175, 184, and 197 MHz, when the incident plane wave propagates  normally to the cylindrical axis. The peak of $Q_b$ at 116~MHz can be correlated to the bandgap \circled{6} for path
$XM$ in Fig.~\ref{PC_circle}(a). Likewise, the peaks at 106, 146, and 165~MHz 
can be respectively correlated to the bandgaps \circled{1}, \circled{4}, and
\circled{6} in Fig.~\ref{PC_circle}(b). However, we were unable to correlate all bandgaps in
Figs.~\ref{PC_circle}(a,b) to the peaks in the spectrum of $Q_b$,
 which indicates that the scattering response of a solitary unit cell is of limited usefulness
in explaining the bandgaps of a phononic crystal.

In  the four band diagrams shown in Fig.~\ref{PC_circle}, we observe that the  group speed $v_g \equiv  d\omega/dk_F  \sim0$  on certain bands.
A zero group speed   indicates that there is no energy flow.
 For a specific path in the IBZ, such eigenfrequencies often fall on a horizontal or a quasi-horizontal  band.
 A \textit{flat band} (for which $v_g     \simeq0$)   arises due to a local resonance, i.e.,
the resonance of an isolated unit cell \cite{local_res1,local_res2}. 
We observe that  some, but not all, bandgaps  lie immediately above and/or below  a flat band in
Figs.~\ref{PC_circle}(a)--(d).  Only a bandgap that lies above and/or below a flat band can be correlated to a local resonance. That is the reason why the bandgap
\circled{6} in Fig.~\ref{PC_circle}(a) and the bandgaps \circled{1}, \circled{4}, and
\circled{6} in Fig.~\ref{PC_circle}(b) could be correlated to the peaks of $Q_b$ identified
in the predecessor study  \cite[Fig.~8(a)]{Chan_AS_JAP}, but  the bandgaps
\circled{1}--\circled{5} in Fig.~\ref{PC_circle}(a) and the bandgaps \circled{2}, \circled{3}, and
\circled{5} in Fig.~\ref{PC_circle}(b) could not be similarly correlated.

 Let us also note that the local resonances underlying the flat bands cannot be
due to absorption. This is because all materials have been taken to be nondissipative,
as is clear from the first paragraph of Sec.~\ref{RD}.

%%%%%%%% Figure 6 begins\ %%%%%%%%%%%%%%%%%%%%%
%%%%%%%% Figure 6 ends %%%%%%%%%%%%%%%%%%%%%%

%%%%%%%% Figure 7 begins\ %%%%%%%%%%%%%%%%%%%%%
	\begin{figure}[H]
			\begin{center}
				\includegraphics[width=3.4in]{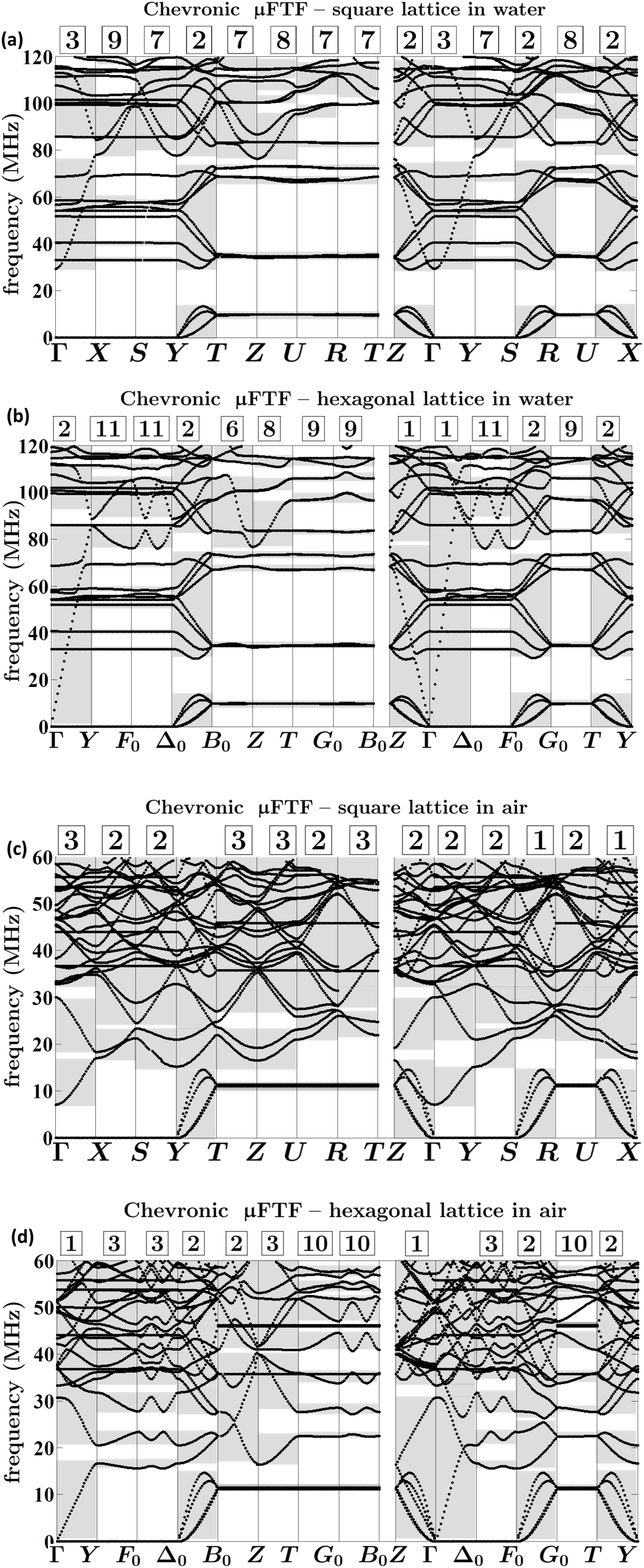}
			\caption{(Color online) Band diagrams for the chosen chevronic $\MFTF$, with the chevronic microfibers
			of Parylene~C arranged on either (a,c) a square or  (b,d) a hexagonal lattice,
			the host medium being either (a,b) water or (c,d) air. Bandgaps are shown unshaded. The boxed number above each path of the IBZ is the number of bandgaps observed on that  path. }
			\label{PC_chevron}
			\end{center}
			\end{figure}
%%%%%%%% Figure 7 ends %%%%%%%%%%%%%%%%%%%%%%

\subsection{Chevronic $\MFTF$}\label{res-chev}

The IBZ of a 3D  unit cell for chevronic $\MFTF$s has a total of 12 paths, whether
the lattice is square or hexagonal.  The band diagrams for the chosen
chevronic $\MFTF$ with the host medium being either water or   air are shown in Figs.~\ref{PC_chevron}(a)--(d). 

When the host medium is water  there is a complete bandgap (13.4--29.7~MHz) for the square lattice; see Fig.~\ref{PC_chevron}(b). This complete bandgap is made of 12 partial
bandgaps. In other words, the chevronic $\MFTF$
is the ultimate bandstop filter, because it does not allow transmission for any incidence direction in that spectral regime, provided that it is sufficiently large in all directions.
 For many, but not all, paths, the complete bandgap widens to 10.1--34.7~MHz. No other complete
bandgap is evident in Fig.~\ref{PC_chevron}.

The  host medium has two distinct effects on the  bandgaps, as can be gleaned by comparing
 Figs.~\ref{PC_chevron}(a,b) with  Figs.~\ref{PC_chevron}(c,d). 
The first effect is on the extent of the spectral regime containing the bandgaps. 
When the host medium is water, bandgaps are found in the $[0.02, 114.7]$-MHz spectral regime, whether the lattice is square or hexagonal. 
That regime narrows down to $[0.02, 40.9]$~MHz and $[0.02, 57.7]$~MHz for the square and the hexagonal lattices, respectively, when the host medium is air. 
The second effect is on the number of partial bandgaps: 58 and 67 partial bandgaps are present for the square and hexagonal lattices, respectively, when the host is water, whereas only 24 and 39 partial bandgaps were identified for the same lattices when the host is  air.
Thus, for the chevronic $\MFTF$, both the number of partial bandgaps and the extent of the spectral regime containing the bandgaps are halved when the host medium is changed from water to air.

Regardless of the host medium, the  chevronic $\MFTF$ has many more partial bandgaps for many symmetric paths of the IBZ than the columnar $\MFTF$. Thus, the former is much more  suitable than the
latter as a multiple-bandstop filter.

In a  predecessor study  on the planewave-scattering characteristics of a single
 chev\-ron  of Parylene~C  in water \cite[Fig.~8(c)]{Chan_AS_JAP}, several peaks in the spectrum of $Q_b$ were found. The spectral locations of these peaks depend on the direction of propagation of the incident plane wave. We found that a partial bandgap (77.2--85.7~MHz) for the path $\Gamma{X}$   in Fig.~\ref{PC_chevron}(a) can be correlated
 to a $Q_b$-peak at 82 MHz, a partial bandgap (76.9--85.8~MHz) for the path $\Gamma{Y}$   in Fig.~\ref{PC_chevron}(a) can be correlated
 to a $Q_b$-peak at 82 MHz, and a partial bandgap (115.7--119.8~MHz) for the path $\Gamma{Y}$   in Fig.~\ref{PC_chevron}(a) can be correlated
 to a $Q_b$-peak at 118 MHz, but no other correlation was found for the square lattice. A similar exercise for the hexagonal lattice was infructuous,
  thereby reaffirming that the resonances of a solitary unit cell can explain some but not all  of the bandgaps of a phononic crystal.
 
Let us also note that the band diagrams in Fig.~\ref{PC_chevron} do not change if the
 chevronic $\MFTF$ is rotated about the $z$ axis by $90^\circ$, i.e., when the chevrons lie in the $yz$ plane instead of the $xz$ plane.

%%%%%%%% Figure 8 begins\ %%%%%%%%%%%%%%%%%%%%%
	\begin{figure}
			\begin{center}
				\includegraphics[width=3.4in]{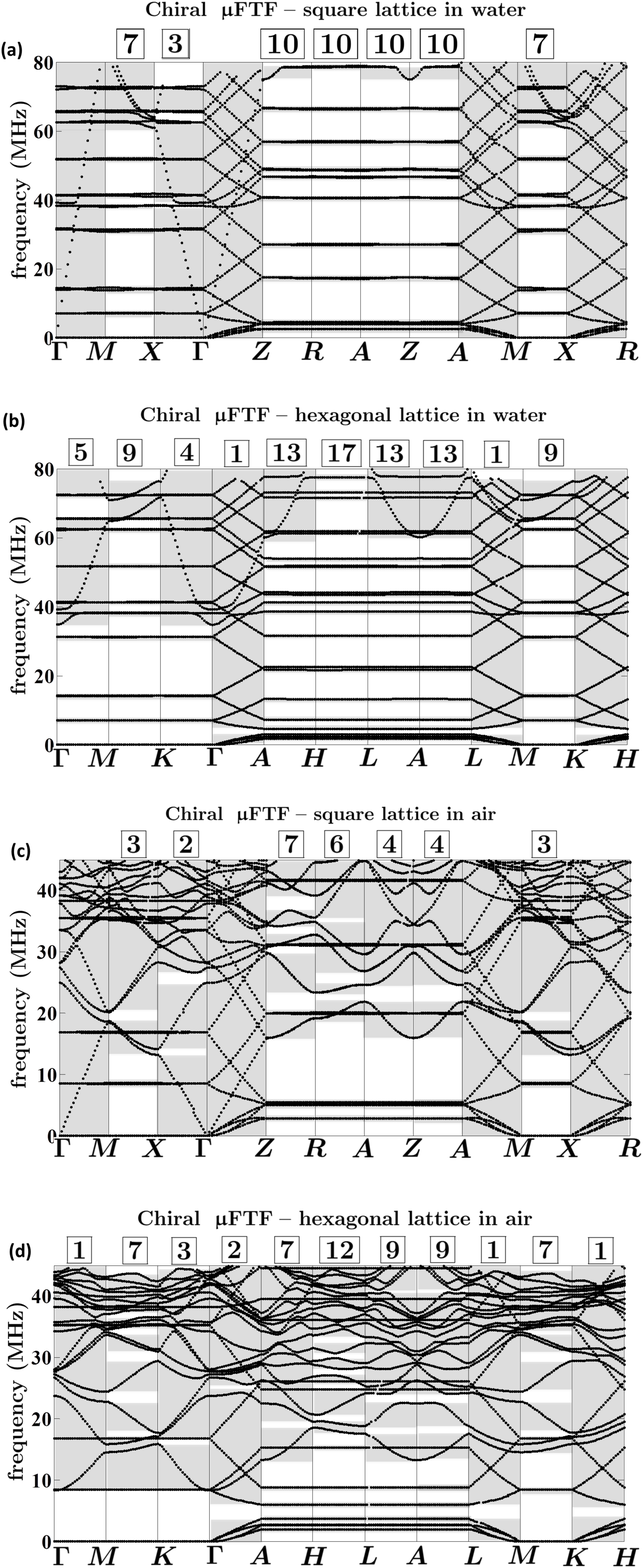}
			\caption{(Color online) Band diagrams for the chosen chiral $\MFTF$, with the upright helical microfibers
			of Parylene~C arranged on either (a,c) a square or  (b,d) a hexagonal lattice,
			the host medium being either (a,b) water or (c,d) air. Bandgaps are shown unshaded.  The boxed number above each path of the IBZ is the number of bandgaps observed on that  path.}
			\label{PC_chiral}
			\end{center}
			\end{figure}
%%%%%%%% Figure 8 ends %%%%%%%%%%%%%%%%%%%%%%

\subsection{Chiral $\MFTF$}\label{res-chi}

The IBZ of a 3D  unit cell for chiral $\MFTF$s has a total of 9 paths, whether
the lattice is square or hexagonal.  The band diagrams for the chosen
chiral $\MFTF$ with the host medium being either water or   air are shown in Figs.~\ref{PC_chiral}(a)--(d). 

When the host medium is water,  there is a complete bandgap  (3.1--4.6~MHz) for the hexagonal lattice, as is evident from Fig.~\ref{PC_chiral}(b).  This bandgap comprises 9 partial bangaps. For four paths of the IBZ, this bandgap widens to 0.01--7.1~MHz. 
When the host medium is air,  there is a complete bandgap (3.7--6.0~MHz) for the hexagonal lattice, as is evident from Fig.~\ref{PC_chiral}(d).  This bandgap comprises 9 partial bangaps. For four paths of the IBZ, this bandgap widens to 0.01--8.5~MHz. No other complete bandgap is evident in Fig.~\ref{PC_chiral}.

Just as for the chevronic $\MFTF$ in Sec.~\ref{res-chev}, the choice of the host medium has two distinct effects on the bandgaps of the chosen chiral $\MFTF$, as becomes clear on comparing
 Figs.~\ref{PC_chiral}(a,b) with  Figs.~\ref{PC_chiral}(c,d). First, when the host medium is water, bandgaps are found in the  0.01--78.6-MHz and 0.01--77.6-MHz spectral regimes for the square lattice and the hexagonal lattice, respectively. The two regimes narrow down to 0.01--39.3~MHz and 0.01--41.3~MHz, respectively, when the host medium is air. Second, 40 and 63 partial bandgaps are present for the square and hexagonal lattices, respectively, when the host is water, whereas only 22 and 43 partial bandgaps were identified for the same lattices when the host is  air.
Thus, for the chiral $\MFTF$---just as for the chevronic $\MFTF$---both the number of partial bandgaps and the extent of the spectral regime containing the bandgaps are halved when the host medium is changed from water to air.

The chiral $\MFTF$ offers many more partial bandgaps for many  than the columnar $\MFTF$
of Sec.~\ref{res-col}. Thus, both the chiral and the  chevronic $\MFTF$s are more  suitable than columnar $\MFTF$ as  multiple-bandstop filters. 

In  Fig.~\ref{PC_chiral}, some but not all bandgaps  lie immediately above and/or below   a flat band and can therefore be ascribed to the resonances of a solitary unit cell.  Comparing Figs.~\ref{PC_chevron} and~\ref{PC_chiral}, we note that the spectral widths of bandgaps are the same for chiral and chevronic $\MFTF$s when the host medium is air, but the spectral widths are lower for the chiral $\MFTF$ than for the chevronic $\MFTF$ when the host medium is water.

Several peaks in the spectrum of the back-scattering efficiency $Q_b$ of a
single-turn helical microfiber of Parylene~C  in water
were found in a  predecessor study \cite[Fig.~8(d)]{Chan_AS_JAP}.
We found for the path $MX$ in Fig.~\ref{PC_chevron}(a) a partial bandgap  (41.6--52.1~MHz)  can be correlated to the $Q_b$-peak at 45 MHz  and another 
partial bandgap  (52.1--61.0~MHz)  can be correlated to  $Q_b$-peaks at 59 MHz and  $61$~MHz. No other correlations between the $Q_b$-peaks and the bandgaps
were found for either of the two  lattices, confirming that the resonances in the scattering response of a solitary unit cell are of limited usefulness
in explaining all of the bandgaps of a phononic crystal.

If all structurally right-handed helical microfibers were to be replaced by their structurally
left-handed counterparts, the Brillouin zone and hence the IBZ would remain unchanged \cite[Fig.~1]{Toader}.  Hence, the eigenfrequencies and the bandgaps would be the same,
regardless of the structural handedness of the chiral $\MFTF$.

%%%%%%%% Figure 9 begins %%%%%%%%%%%%%%%%%%%%%
	\begin{figure}
			\begin{center}
				\includegraphics[width=5in]{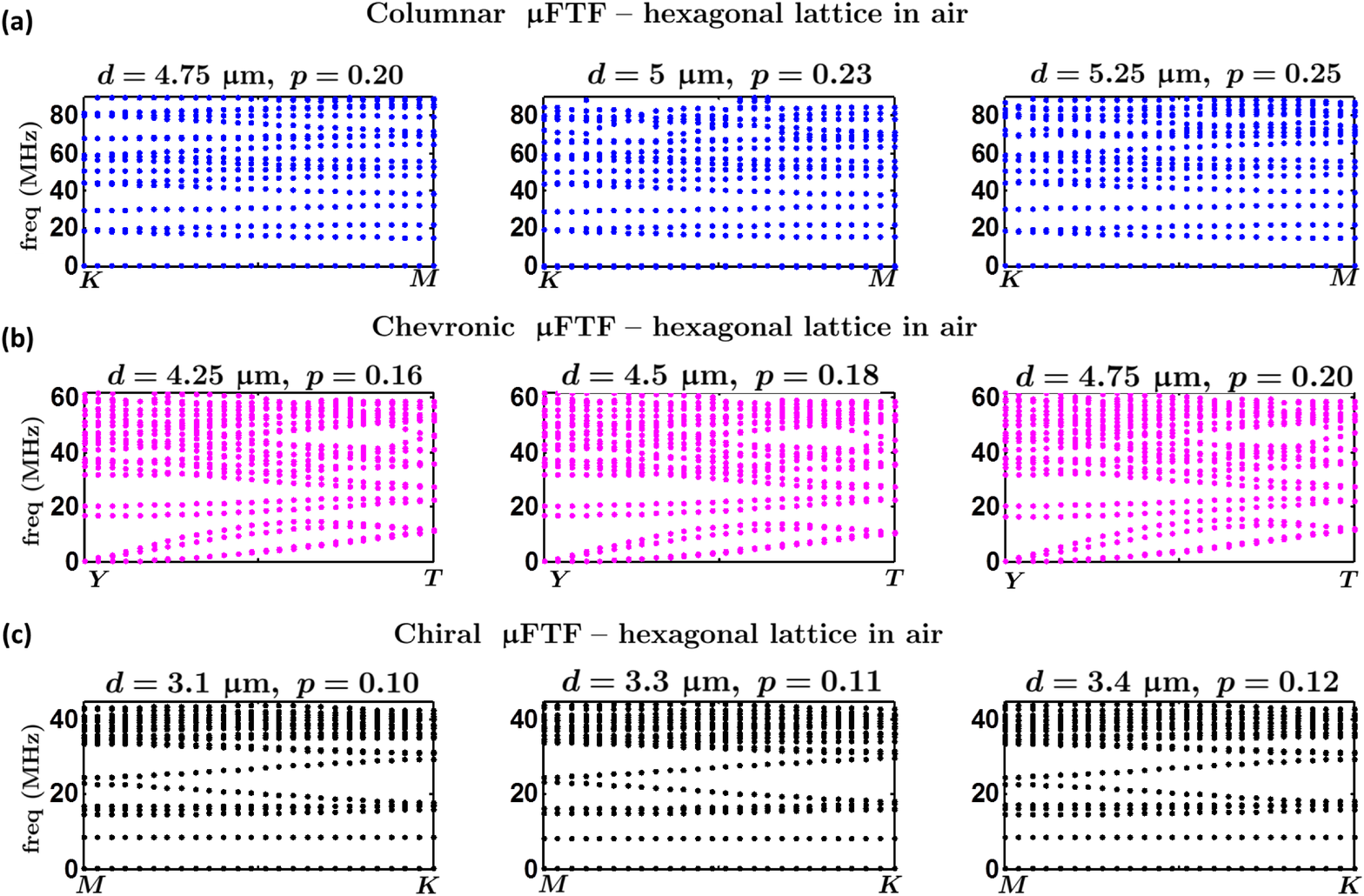}
			\caption{(Color online)  Band diagrams  (a) on path
			$KM$ for  columnar $\MFTF$s,  (b) on path $YT$ for chevronic $\MFTF$s, and (c) on path $MK$ for chiral $\MFTF$s, when the lattice is hexagonal and the host medium is air.  Values of $d$ and $p$ are marked for each band diagram,
whereas $a$, $b$, and $\beta$ (as applicable) were
left unaltered from their values stated in Fig.~\ref{Unit_cells}.
		}
			\label{ffeffect}
			\end{center}
			\end{figure}
%%%%%%%% Figure 9 ends %%%%%%%%%%%%%%%%%%%%%% 

\subsection{Stability of band diagrams}\label{stab}
Several flat bands exist in the band diagrams shown in Figs.~\ref{PC_circle}, \ref{PC_chevron}, and \ref{PC_chiral}, each flat band being indicating of a local resonance
\cite{local_res1,local_res2}.
Among the four choices for the combination of the lattice and the host medium, the phononic crystal comprising microfibers arranged in a hexagonal lattice and immersed in air was found to have many  flat bands. Furthermore, the band diagrams contain many partial and few complete bandgaps. Bandgaps are attractive for filtering applications, because of
either appreciable insensitivity (in the case of a partial bandgap) or total   insensitivity (in the case of a complete bandgap) to the direction of propagation of the  plane wave incident on a phononic 
crystal of finite thickness.
As dimensional variations can creep in during fabrication, the stability of
flat bands and bandgaps with respect to several perturbations of the unit cells
were investigated by us.

Let us now present the results of a representative perturbation---that of
the filling fraction $p$ on the band diagrams, the host medium being air.
For this investigation on a specific band diagram, we selected one path of the  IBZ containing a   large number of flat bands, and computed the band diagram for the same path by altering the microfiber diameter $d$ in order to change $p$, whereas $a$, $b$, and $\beta$ (as applicable) were
left unaltered.

 The calculated band diagrams for $p$ varying by about $\pm12\%$
are shown in Fig.~\ref{ffeffect}. 
In each row of this figure, the center panel is the band diagram for the unaltered $p$ (Fig.~\ref{Unit_cells}), whereas the left and right panels are band diagrams for lower
and higher values of $p$, respectively. The eigenfrequencies vary by no more than
$\pm5\%$ for $p$ varying by about $\pm12\%$.
The flat bands  remained virtually unchanged, thereby confirming
that they are   due to  local resonances \cite{local_res1,local_res2}. Also, the bandgaps change by less than $\pm6\%$ for the $p$-variations
considered. We concluded from these and other numerical results that manufacturing variations will not drastically affect the
performances of the phononic crystals under consideration.

\subsection{Discussion}
 
As compared to scatterers of simple shapes  such as circular cylinders \cite{Pennec2010229} and spheres \cite{sphere_PWE}, we deduce from  Figs.~\ref{PC_circle} to \ref{PC_chiral} that the number of partial bandgaps increases as the complexity of the shape of the scatterer in the
unit cell increases.
 The increase in the number of partial bandgaps is not just due to the increase in the number of symmetric paths of the IBZ, but because of  increase in the  number of partial bandgaps per path---as can be established by dividing the number of partial bandgaps by the number of unique paths for the IBZ in a band diagram.
 
From Figs.~\ref{PC_circle} to \ref{PC_chiral}  we note that chiral $\MFTF$s have lower eigenfrequencies compared to columnar and chevronic
$\MFTF$s. Hence,  the center-frequencies of the bandgaps are also lower 
for the chiral $\MFTF$s than for the columnar and chevronic
$\MFTF$s.
 The columnar $\MFTF$ were found to have the highest center-frequencies. 
These observations are in complete agreement with the eigenfrequencies of the individual microfibers found previously by us \cite[Table~II]{Chan_AS_JAP}.

Existing bulk-acoustic-wave (BAW) filters  operate in the higher-MHz and the lower-GHz regimes \cite{BAWF} as narrow-bandstop filters.
Requiring  tedious   fabrication procedures \cite{BAW_fab},  BAW filters are used in 0.3--3-MHz regime in aviation and AM radio circuits, in  the 3--30~MHz regime in shortwave radio circuits, and in the 30--300~MHz regime  in FM radio circuits.
Given their  simple fabrication procedures \cite{LaiWei_12_1_MRI,WLRR,protein_assay}, 
chevronic and chiral $\MFTF$s of Parylene~C can be of use in  aviation,  AM radio, and
shortwave radio circuits as bandstop filters.
Parylene-C  columnar $\MFTF$ can be used as bandstop filters upto 163~MHz in  FM radio circuits. 

Surface-acoustic-wave (SAW) devices often used for mobile and 	wireless communications \cite{Colin_campbell}, operate in the 0.1--1000-MHz regime. These devices are also used in microfluidics \cite{Leslie}. 
 The manufacture of  SAW devices requires a series of steps in a layer-by-layer  procedure, as well as the fabrication of an intricate  pattern of metallic interconnects  \cite{SAW_Patent}.
Given the large number of bandgaps on various paths of the IBZs in Secs.~\ref{res-col}--\ref{res-chi}, we propose that  Parylene-C $\MFTF$s fabricated over piezoelectric substrate can be used as SAW filters.  

Based on Figs.~\ref{PC_circle} to \ref{PC_chiral}, we note that our Parylene microfibrous thin films have bandgaps in the  1--160-MHz regime.
Since ultrasonic transducers operate in the 0.5--25-MHz regime \cite{Ultrasonic_frequency},
Parylene-C $\MFTF$s can be cascaded onto these transducers to filter out specific spectral regimes.
Thus,  Parylene-C $\MFTF$s can be classified as low-frequency ultrasonic  filters and 
thin-film bulk acoustic resonators. We have numerically ascertained that bandgap engineering is possible by changing the filling fraction $p$, but that issue lies outside the scope of this paper.

Parylene~C is more amorphous in its microfibrous form  than in its bulk form \cite{Chan_AppSci}, indicating that  the Parylene-C $\MFTF$s are softer than the bulk Parylene~C. 
Since the $\MFTF$s are softer, the distance between adjacent microfibers can easily be adjusted by the application of an external pressure or strain \cite{FeiWang, Wang20122881}, similar to tunable photonic crystals \cite{Gopalan} and cholesteric elastomers \cite{SKF,Shibaev,Varanytsia}.
As the bandgaps can therefore be dynamically tuned  for every combination of microfiber morphology and   lattice,  Parylene-C $\MFTF$s  can be called \textit{soft tunable phononic crystals}. 
 \\

\section{Concluding Remarks}
Parylene-C  microfibrous thin films     were treated as phononic crystals comprising 
identical  microfibers arranged either on a square or a hexagonal lattice. The microfibers
could be columnar, chevronic, or helical in shape, and  the host medium could
be either  water or air. 

For these $\MFTF$s with microfibers of  realistically chosen dimensions  
\cite{Chan_AS_JAP}, all the  bandgaps were observed to lie in the 0.01--162.9-MHz regime.
Complete bandaps were observed for the following $\MFTF$s: 
(i) columnar $\MFTF$ with microfibers arranged on a hexagonal lattice in air, 
(ii) chevronic $\MFTF$ with microfibers arranged on a square lattice in water, and 
(iii) chiral $\MFTF$ with microfibers arranged on a hexagonal lattice in either water or air. 
The upper limit of the frequency of bandgaps was the highest
for the columnar $\MFTF$s and the lowest for the chiral $\MFTF$s. 
For all  $\MFTF$s  partial  bandgaps along many symmetric directions were found. The
number of partial bandgaps was higher when the host medium is water than air.

The obtained bandgaps for all the Parylene-C $\MFTF$s suggests their possible use as multi-band
bulk-acoustic-wave filters. These filters can be used in conjunction with
 ultrasonic transducers as well as  surface-acoustic-wave devices.
By patterning the substrates a higher inter-microfiber distance could be achieved and the desired bandgaps in the lower MHz regime can be obtained. 
Furthermore, the low elastic modulus of Parylene C also makes the $\MFTF$s suitable for mechanical tuning. We are currently investigating the terahertz photonic properties of 
these engineered micromaterials, {and plan to extend our work to other parylenes
such as Parylene~N \cite{PCN_similar_2}.}\\

\noindent \textbf{Acknowledgments.}
We  thank  the Research and Cyberinfrastructure  Center of Institute of Cyber Science (ICS) of the Pennsylvania State University for  computing resources,
Brian Van Leeuwen and Hirofumi Akamatsu of the Department of Materials Science and Engineering (Penn State) for  help in identifying irreducible Brillouin zones, and  Anand Kumar Singh (ICS)  and Chien Liu (COMSOL support team) for assistance
with mesh-related issues.
CC and AL are grateful to the Charles Godfrey Binder Endowment at
Penn State for  financial supporting this research.

\end{document}